\documentclass[journal=mamobx,manuscript=article]{achemso}

\usepackage[version=3]{mhchem} 
\usepackage{siunitx}
\DeclareSIUnit{\rad}{rad}

\author{John J. Karnes}
 \email{karnes@llnl.gov}
 \affiliation{%
 {Lawrence Livermore National Laboratory
       Livermore, California 94550, United States}
}
\author{Todd H. Weisgraber}%
\affiliation{%
 {Lawrence Livermore National Laboratory
       Livermore, California 94550, United States}
}
\author{Caitlyn C. Cook}
\affiliation{%
 {Lawrence Livermore National Laboratory
       Livermore, California 94550, United States}
}
\author{Daniel N. Wang}
\affiliation{%
 {Lawrence Livermore National Laboratory
       Livermore, California 94550, United States}
}
\author{Jonathan C. Crowhurst}
\affiliation{%
 {Lawrence Livermore National Laboratory
       Livermore, California 94550, United States}
}
\author{Christina A. Fox}
\affiliation{%
 {Lawrence Livermore National Laboratory
       Livermore, California 94550, United States}
}
 \alsoaffiliation{%
 {Department of Materials Science and Engineering, University of California, Davis, Davis, California 95616, United States}
}
\author{Bradley S. Harris}
 \affiliation{%
 {Department of Chemical Engineering, University of California, Davis, Davis, California 95616, United States}
}
\author{James S. Oakdale}
\affiliation{%
 {Lawrence Livermore National Laboratory
       Livermore, California 94550, United States}
}
\author{Roland Faller}
\email{rfaller@ucdavis.edu}
 \affiliation{%
 {Department of Chemical Engineering, University of California, Davis, Davis, California 95616, United States}
}
\author{Maxim Shusteff}
\affiliation{%
 {Lawrence Livermore National Laboratory
       Livermore, California 94550, United States}
}

\title{Isolating Chemical Reaction Mechanism as a Variable with Reactive Coarse-Grained Molecular Dynamics: Step-Growth versus Chain-Growth Polymerization}

\abbreviations{IR,NMR,UV}

\begin{document}

\begin{abstract} 
  We present a general approach to isolate chemical reaction mechanism as an independently controllable variable across chemically distinct systems. Modern approaches to reduce the computational expense of molecular dynamics simulations often group multiple atoms into a single ``coarse-grained'' interaction site, which leads to a loss of chemical resolution. In this work we convert this shortcoming into a feature and use identical coarse-grained models to represent molecules that share non-reactive characteristics but react by different mechanisms.  As a proof of concept we use this approach to simulate and investigate distinct, yet similar, trifunctional isocyanurate resin formulations that polymerize by either chain- or step-growth. Since the underlying molecular mechanics of these models are identical, all emergent differences are a function of the reaction mechanism only. We find that the microscopic morphologies resemble related all-atom simulations and that simulated mechanical testing reasonably agrees with experiment.
\end{abstract}

\section{Introduction}Chemical reaction mechanisms, along with intra- and intermolecular forces and other processing conditions, dictate the formation of macromolecular structures and materials from smaller molecular building blocks. Understanding the contributions of these factors permits their use as design parameters to control the micro- and macroscopic properties of the resulting new material. However, deconvolution of these factors is often not possible since the molecular mechanics of reactive sites and reaction mechanism are intertwined. 

This directs us toward a general question: Can we isolate \textit{chemical reaction mechanism} as an independent variable? This is an open and fundamental question in chemical physics. While different reaction mechanisms may be compared experimentally~\cite{yokoyamaConvertingStepGrowthChainGrowth2007,asakawaCationicPolymerizationPhenyl2020,cookHighlyTunableThiolEne2020}, these approaches require unavoidable approximations. Computer simulations do not necessarily share this limitation and can directly address this thought experiment. In this work, we introduce a new approach that converts a drawback of coarse-grained computer simulations into a key feature that enables direct comparison of chemically distinct systems, isolating the chemical reaction mechanism as an independent variable.

In conventional molecular dynamics (MD) simulations each atom is represented by only a few parameters and equations that reproduce the intra- and intermolecular interaction energies for a given system. Integration of Newton’s equations of motion then evolves this system over time. All-atom models and the resulting simulations naturally mesh with physical and chemical intuition. However, even with increases in computational power, all-atom classical MD is still limited in length scale and simulation duration. To loosely quantify, simulating several 10 nm$^3$ boxes of $10^5$ atoms, each for hundreds of nanoseconds, taxes the resources of most researchers. These limitations become more pronounced for high molecular weight molecules that require both larger simulation boxes and longer simulation times to capture the correct interactions and dynamics.~\cite{kremerCrossoverRouseReptation1988,auhlEquilibrationLongChain2003} 

Coarse-grained MD (CGMD) approaches that combine multiple atoms into a single representative “bead” are a popular approach toward reducing computational expense. Calculations are accelerated by reducing the number of interaction sites and allowing larger time steps by removing the fastest motions.~\cite{vothCoarseGrainingCondensedPhase2008} Such techniques have brought on valuable insight into many fields of soft matter physics, e.g. entangled dynamics in polymer melts, impact of curvature and surfaces on biomimetic membranes, etc.~\cite{kremerDynamicsEntangledLinear1990,everaersKremerGrestModels2020,reynwarAggregationVesiculationMembrane2007,xingInteractionsLipidBilayers2008}
Recently CGMD has been extended to reactive systems for a range of systems, including the simulated polymerization of poystyrene,~\cite{farahReactiveMolecularDynamics2010} polyurethane,~\cite{ghermezcheshmeMARTINIbasedSimulationMethod2019} and silica.~\cite{carvalhoStickyMARTINIReactiveCoarsegrained2022} Notably, CGMD polymerization approaches have been able to capture characteristic features of the resulting materials, reproducing thermomechanical and morphological properties like the self-assembly into micellar structures or amorphous clusters, depending on simulation conditions.~\cite{ghermezcheshmeMARTINIbasedSimulationMethod2019,carvalhoStickyMARTINIReactiveCoarsegrained2022}  

When atoms are merged or `coarse-grained' into a single representative site, the potential energy landscape becomes smoother, more simplified.~\cite{muller-platheCoarseGrainingPolymerSimulation2002,meinelLossMolecularRoughness2020} The identity of reactive functional groups may disappear when merging individual atoms into beads. We may refer to this as a `loss of chemical resolution.' In this work, we take advantage of both CGMD’s ability to simulate crosslinked polymer networks and the accompanying loss of chemical resolution. We represent chemically distinct monomers with similar intramolecular structures but different reactive sites with the same coarse-grained model: a `universal monomer' (UM). Despite the different monomers being represented by precisely the same model, the UM, we preserve the respective reaction mechanisms native to each monomer species. In practice, we simulate a box of the `universal monomer' and impart chemical identity by implementing the chemical reaction mechanism as a modular `rule,' allow the selected reaction to proceed, and analyze the resulting material. We then repeat with a new liquid box and different reaction mechanism. Since the molecular interactions of the monomer models are precisely the same, all differences in the evolution of the microscopic network and properties of the resulting materials emerge entirely from the different reaction mechanisms.

This approach toward isolating reaction mechanism as a variable borrows elements from cellular automata (CA), as defined by Rucker, where a discrete set of elements are each in a `state' and these elements' states are updated in parallel using a homogenous and local `rule'.~\cite{ruckerContinuousvaluedCellularAutomata2003} In our simulations the UM's reactive sites may be considered as elements in binary on/off (reacted/unreacted) states and the reaction mechanisms are analogous to CA rules. There is not a large formalistic distance between these reactive CGMD simulations and CA simulations of related physical systems~\cite{dostrovsky50MechanismSubstitution1946,frenkelLatticeGasesPolymers1990,smithCellularAutomatonSimulation1991} and we regard the present work as a new approach toward designing reactive CGMD that incorporates the underlying philosophies of CA and exhibits the same interesting emergent behaviors.~\cite{ruckerContinuousvaluedCellularAutomata2003,gardenerMathematicalGamesFantastic1970,wolframStatisticalMechanicsCellular1983,ostrovContinuousvaluedCellularAutomata1996} By implementing the reaction mechanism as a `rule,' we may rapidly quantify the resulting polymer's micro- and macroscopic properties as a function of polymerization mechanism alone. 

The rapidly evolving additive manufacturing (AM) landscape provides a recent example where reaction mechanism is used as a design variable.  Compare Volumetric Additive Manufacturing (VAM)~\cite{shusteffOnestepVolumetricAdditive2017,kellyVolumetricAdditiveManufacturing2019,loterieHighresolutionTomographicVolumetric2020} with two-photon polymerization (2PP).~\cite{miwaFemtosecondTwophotonStereolithography2001,zhouReviewProcessingAccuracy2015} Both approaches accomplish three-dimensional (3D) printed structures via photopolymerization. Whereas 2PP sequentially scans a tightly-focused sub-micron laser focal volume to write a structure in 3D space, VAM concurrently solidifies all points within a 3D object by illuminating a rotating volume of photosensitive resin with a dynamically evolving light pattern.\cite{kellyVolumetricAdditiveManufacturing2019} These techniques use vastly different irradiation times: a few nanoseconds of femtosecond pulses in 2PP and minutes of illumination in VAM. Techniques like 2PP are enabled by the fast reaction kinetics of multi-functional acrylate resins, and nascent techniques like VAM naturally borrow feedstocks from more mature AM technologies like 2PP during the development phase. However, the longer irradiation time of VAM allows using resins with slow kinetics, thus enabling exploration of new chemistries. Since reaction mechanism profoundly effects the printed part's material properties, chemical reaction mechanism becomes a design parameter.~\cite{cookHighlyTunableThiolEne2020} In this work, Cook et. al. successfully replaced the pendant acrylate groups in their photoresin with thiol and alkene functionalities, effectively switching the reaction mechanism within their VAM instrument from radical-initiated chain-growth to a radical step-growth `click' chemistry.~\cite{cookHighlyTunableThiolEne2020} Swapping the functional groups thus resulted in more mechanically-robust printed parts by photopolymerization-based additive manufacturing. 

We select this use case as proof-of-concept to demonstrate our new computational approach and show that computer simulations are well-positioned to rapidly explore this design space and avoid lengthy and costly \textit{formulate-cure-test} development cycles. Our work begins with the same three trifunctional isocyanurate monomers used by Cook et al. and uses the UM approach to isolate differences that emerge from changes in reaction mechanism and resulting microscopic network topologies. This isolation is possible since the UM approach ignores differences in the monomers' molecular mechanics introduced by the different functional groups.

Many contemporary polymer simulations focus on a specific polymer or class of polymers,~\cite{luoEffectLoopDefects2020,torres-knoopModelingFreeradicalPolymerization2018,torres-knoopEffectDifferentMonomer2021,gaoEffectEpoxyMonomer2017,karnesNetworkTopologyCrossLinked2020} and we use similar analytical approaches to investigate and validate the work presented here. However, the methodology we introduce with the UM approach is more closely related to studies where researchers systematically disabled components of MD force fields. For example, one sets partial charges of a molecule or chemical moiety to zero to quantify the contribution of electrostatic interactions to local intermolecular ordering, structure, and dynamics~\cite{karnesLocalIntermolecularOrdering2017,karnesNetworkTopologyCrossLinked2020}. Gissinger and co-workers disabled chain-chain linking in reactive all-atom MD simulations of styrene, resulting in polystyrenes with polydispersity and cyclic structures unlike that of the experimental system.~\cite{gissingerREACTERHeuristicMethod2020} Although nonphysical, this \textit{in silico} deconstruction reveals the contributions of force field components to emergent physical and chemical properties.

In the remainder of this work we provide a detailed overview of the UM approach but emphasize that this method is transferable and not reliant on any particular coarse-graining strategy or implementation of molecular dynamics. This is followed by analysis and discussion of proof-of-concept simulations, conclusions and thoughts on future work.

\section{Computational methods}

\subsection{`Universal monomer' design and parameterization}

\begin{figure}[h]
\includegraphics{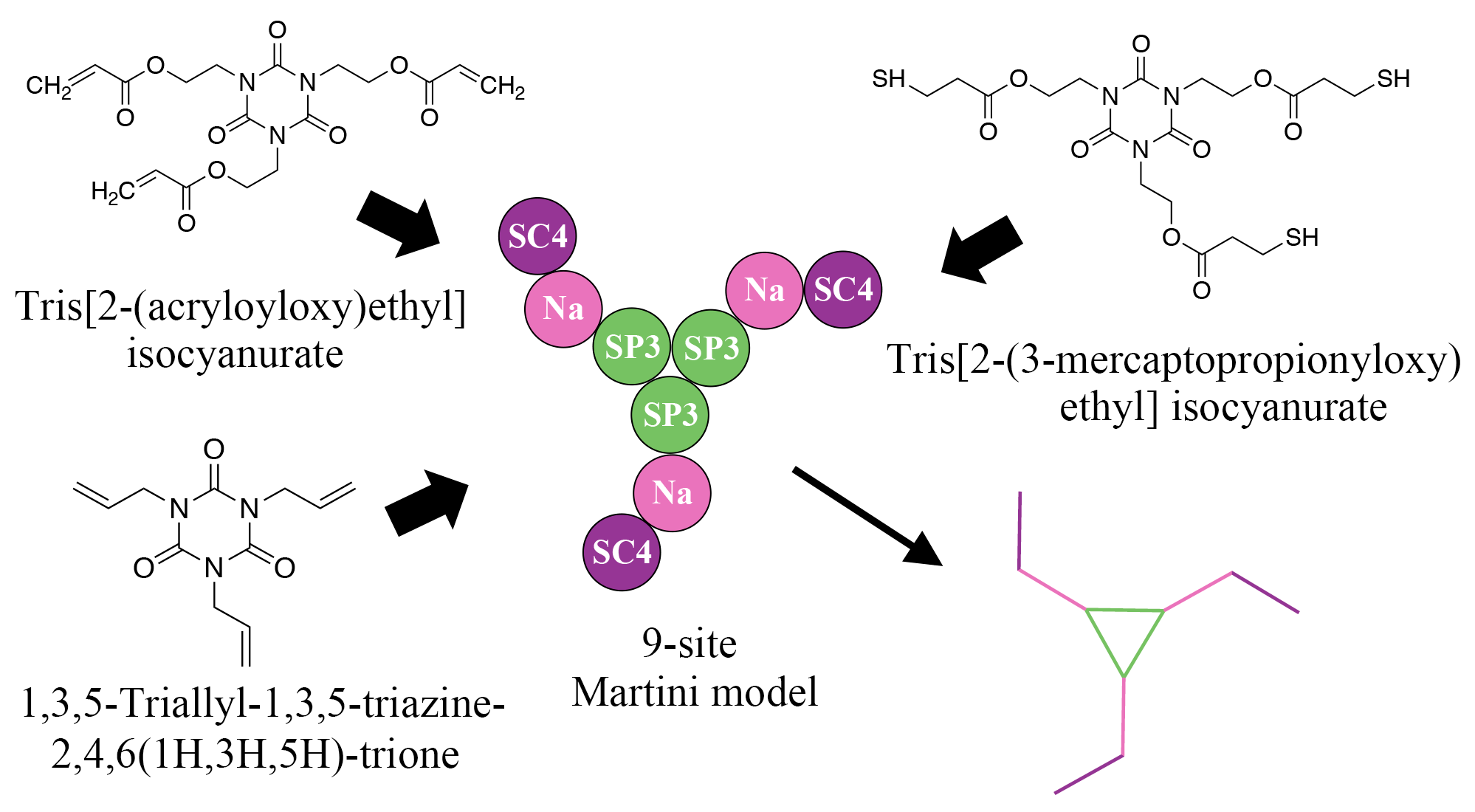}
\caption{\label{fig:um} Trifunctional acrylate and thiol-ene isocyanurate monomers are represented by a 9-site Martini model. Bead labels in the central image are consistent with types from Reference \citenum{marrinkMARTINIForceField2007}. Lower right: simple stick model of the ‘universal monomer.’}
\end{figure}

In this proof-of-principle work, our `universal monomer' uses the functional form and intermolecular potentials of the Martini 2 force field.~\cite{marrinkMARTINIForceField2007} In the Martini coarse-graining approach, typically 4 heavy atoms and their associated hydrogens are merged into a single interaction center or ‘bead.’ For example, the 4-carbon butane molecule is represented by one ‘alkane’ bead, octane by two.~\cite{marrinkMARTINIForceField2007} The interatomic (`inter-bead') potential is represented as the pairwise sum of the Lennard-Jones potential 
\begin{equation}
  U_{ij}(r) = 4\epsilon_{ij}\left[\left(\frac{\sigma_{ij}}{r}\right)^{12}-\left(\frac{\sigma_{ij}}{r}\right)^{6}\right],
\label{eqn:LJ}
\end{equation}
where $i$ and $j$ are beads separated by a distance $r$. Intramolecular potentials consist of harmonic bonds and angles, 
\begin{equation}
  V_{\text{bond}}(r) = \frac{1}{2}k_\text{bond}(r-r_{eq})^2,
\label{eqn:bond}
\end{equation}
\begin{equation}
  V_{\text{angle}}(\theta) = \frac{1}{2}k_\text{angle}(\text{cos}(\theta)-\text{cos}(\theta_{eq}))^2,
\label{eqn:angle}
\end{equation}
and force constants $k$ which we reparametrized from standard Martini values to better reflect the average geometry of the isocyanurates. Additionally, we implemented an improper dihedral potential of the form
\begin{equation}
  V_{\text{imp}}(\theta) = k_\text{imp}(\theta-\theta_{eq})^2
\label{eqn:imp}
\end{equation}
to enforce the planar geometries of the monomers. While our parameterization approximates the molecular mechanics of these isocyanurates, we emphasize that the chemical identity of the monomer’s reactive site, i.e. acrylate, thiol, or ene, is buried within a single Martini bead. Figure \ref{fig:um} shows the three trifunctional isocyanurate monomers used in this work and the 9-site ‘universal monomer’ developed to represent all three molecules. 

Custom intramolecular parameters for the 9-site universal monomer were derived from all-atom simulations of the three trifunctional isocyanurates (ICN). To obtain this all-atom data three simulation boxes, one for each isocyanurate species, were prepared. Each box contained 512 copies of the respective monomer and each monomer was modeled using the OPLS-AA force field~\cite{jorgensenPotentialEnergyFunctions2005,dodda14CM1ALBCCLocalized2017,doddaLigParGenWebServer2017}. Each of the boxes was relaxed for 20 ns of simulation time followed by a 10 ns production run. 
These all-atom simulations used a timestep of \SI{1}{fs} and implemented the Nos\'e--Hoover thermostat and barostat with damping parameters of 100 and 1000 timesteps, respectively.~\cite{noseMolecularDynamicsMethod1984,noseUnifiedFormulationConstant1984,hooverCanonicalDynamicsEquilibrium1985} Long-range Coulombic interactions were calculated using the particle-particle particle-mesh solver.~\cite{hockneyComputerSimulationUsing1999} From each production run trajectory 10\textsuperscript{6} configurations were stored for analysis.

Individual atoms were assigned to Martini beads,~\cite{marrinkMARTINIForceField2007} a detailed breakdown is included as Figure S1 in the Supporting Information. We note that the isocyanurates' pendant reactant arms are of different length, as seen in Figure \ref{fig:um}. This aspect of the UM parameterization is an unavoidable concession in this proof-of-principle work, and we chose to include the ``short-armed'' tri-ene (triallyl isocyanurate) used in related experimental studies~\cite{mongkhontreeratUVInitiatedThiol2013,cookHighlyTunableThiolEne2020} versus replacing the tri-ene with a more morphologically similar tri-ene (e.g. insertion of a longer alkyl tether), which results in a realistic test of this new approach. We reserve a detailed deconstruction of UM parameterization and extension to combinations of bi- and tri- functional monomer formulations for future work.

Using in-house code, we calculated coarse-grained bonds, bends, and SP3-SP3-SP3-Na improper dihedrals based on the centers of mass of each bead grouping. These CG-mapped geometries were collected for each monomer in each saved configuration and assembled into probability distributions representing each component of the coarse-grained force field topology.

For parameterization, our initial coarse-grained simulation used the Martini 2 force field~\cite{marrinkMARTINIForceField2007} without modification and was performed using a similar approach: 512 copies of the 9-site model. Similar probability distributions representing the coarse-grained intramolecular geometries were obtained from this `standard' Martini force field. We used a simple iterative approach to optimize the intramolecular geometry so that it provided a reasonable description of all three of the trifunctional monomers. A full collection of the Martini, OPLS-AA, and UM probability distributions is collected as Figure S2 in the Supporting information. Intramolecular force field parameters are summarized in Table~\ref{table:fftable}. As mentioned earlier, intermolecular potentials were taken from the Martini 2 force field and used without modification.

\begin{table}[ht!]
\centering
\begin{tabular}{ p{3.3cm} c  c }
 \hline
 bond & $r_{eq}$(\AA) & $k_{eq}$(kcal/mol)\\
 \hline
 SC4-Na & 3.50 & 2.70\\
 Na-SP3 & 3.50 & 2.10\\
 SP3-SP3 & 2.93 & 23.9\\
 SC4-SC4 & 4.30 & 23.9\\
 \hline
 \hline
  bend & $\theta_{eq}$(deg) & $k_{eq}$(kcal/mol)\\
 \hline
 SP3-Na-SC4 & 180 & 2.99\\
 SP3-SP3-Na & 180 & 2.99\\
 \hline
 \hline
  improper & $\phi_{eq}$(deg) & $k_{eq}$(kcal/mol)\\
 \hline
 SP3-SP3-SP3-Na & 20 & 16.0\\
 \hline
 \end{tabular}
 \caption{Intramolecular force field parameters used by the `universal monomer.'}
\label{table:fftable}
\end{table}

\clearpage

\subsection{Reactive CGMD simulation details}

All production run simulations began with an equilibrated liquid box of 2000 UMs. At the start of the reactive trajectory we decide to simulate either an acrylate (radical chain-growth) or thiol-ene (radical step-growth) system, it is precisely at this point that we recast the loss of chemical resolution imparted by coarse-graining into an advantage. Chain-growth polymerization is simulated by defining all 2000 UMs to be triacrylates. Similarly, a step-growth system may be created by randomly defining 1000 UMs to be tri-thiols and the remaining 1000 to be tri-enes. After assigning chemical identity to the UMs the reactive trajectory may begin, with polymerization progressing by the selected mechanism. 

\begin{figure}[h]
\includegraphics[scale=0.65]{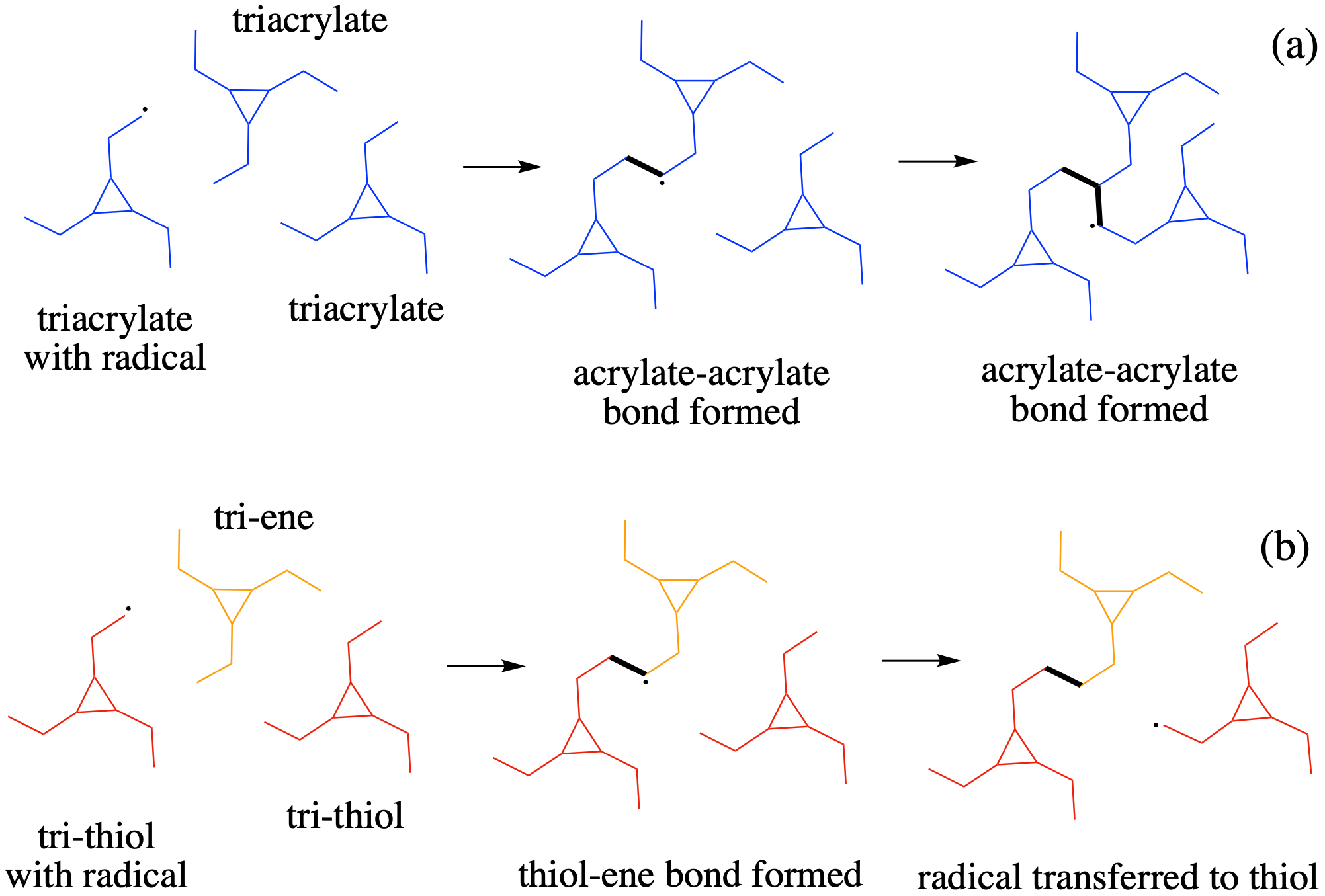}
\caption{\label{fig:mech} Top row: In  acrylate radical chain-growth, the radical moves along the propagating chain as more acrylates are added. Bottom row: In thiol-ene radical step-growth, the reactive moieties only participate in one new bond. After the thiol-ene bond forms, the radical is transferred to a nearby, unreacted thiol group.}
\end{figure}

Cartoons in Figure \ref{fig:mech} graphically summarize the two reactive schemes implemented in this work. In the acrylate system, a radical-bearing acrylate site $i$ may form a new bond with an unreacted radical site $j$. Upon formation of this new bond, the radical is transferred to site $j$, which may then react with other unreacted acrylate sites, forming extended $i-j-k-l-m...$ chains of linked acrylate beads. In the thiol-ene system a radical-bearing thiol site may form a new bond with an unreacted ene site, with  the radical then transferred to the ene site. This radical ene does not subsequently form new covalent bonds as in the acrylate system. The radical on the ene site is instead transferred to a nearby unreacted thiol site and the cycle continues. This results in the formation of single monomer-monomer linkages in the thiol-ene system. We use a distance-based cutoff approach to simulate both radical chain- and step-growth polymerization in the LAMMPS molecular dynamics environment.~\cite{gissingerModelingChemicalReactions2017,thompsonLAMMPSFlexibleSimulation2021,plimptonFastParallelAlgorithms1995} 

All data shown represents the average of 6 independently-prepared simulations for each reactive mechanism. Analyses were performed using in house code and the NetworkX Python package.~\cite{hagbergExploringNetworkStructure2008} 

\section{Results and Discussion}

We track the progress of each polymerization reaction as functional group conversion ($X$), the percentage of theoretically possible new monomer-monomer linkages that formed during the reactive trajectory. The quantification of $X$ in the two systems differs and is important to clarify: In the ‘acrylate’ chain-growth system, each acrylate functional group may participate in two acrylate-acrylate bonds. A single polymerization reaction in a propagating chain creates one new bond for each bonding partner, thus consuming one net acrylate vinyl group. Therefore, in a simulation of 2000 trifunctional monomers, $X = 100\%$ involves converting $3 \times 2000 = 6000$ vinyl groups, forming 6000 new covalent bonds. In the ‘thiol-ene’ radical step-growth mechanism, each new bond fully consumes two functional groups: one thiol and one ene group, each only able to form one new covalent bond through thiol-ene polymerization. Therfore $X = 100\%$ corresponds to 3000 new bonds in the thiol-ene step-growth simulations. For a given conversion, the chain-growth system has twice the number of new bonds as the equivalent step-growth system.

\begin{figure}[h]
\includegraphics[width=3.3in]{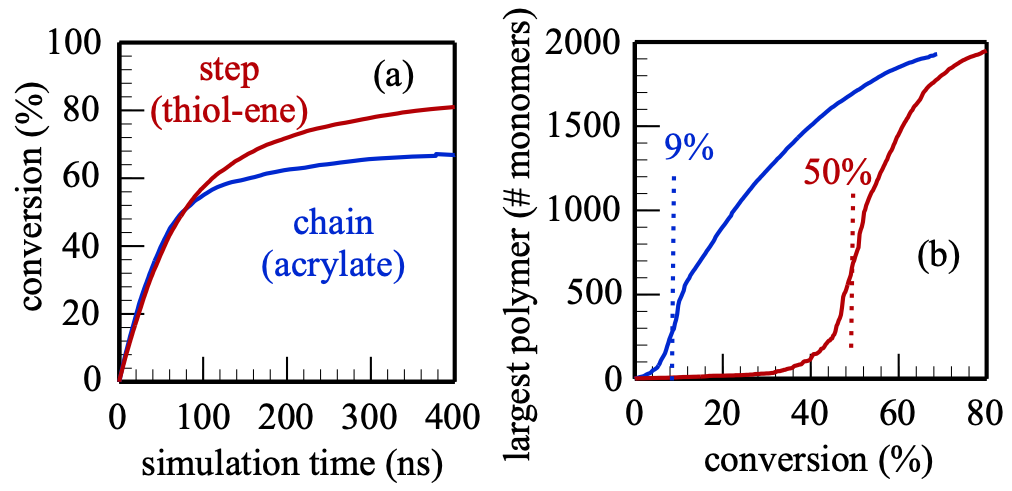}
\caption{\label{fig:react} (a) Conversion versus simulation time. (b) Size of the largest propagating polymer unit as a function of conversion for the chain- and step-growth systems. Vertical dashed lines in (b) guide the eye to the inflection points to $X_{\text{gel}}$.}
\end{figure}

Figure~\ref{fig:react}a shows conversion versus simulation time for each system. In these simulations both systems begin reacting at approximately the same rate. The reaction rates also begin to slow at approximately the same point, around 75 ns of simulation time, with a more noticeable rate reduction and ultimately lower endpoint conversion in the acrylate system than in the thiol-ene system. While these reactive schemes simplify the kinetics and energetics of their experimental analogues, they reasonably reflect experimental endpoint conversions and the relative differences. For comparison, fully cured samples of the same monomer species have endpoint conversions of $X=66\%$ and $X=89\%$ for the acrylate and tholene systems, respectively, versus $X=68\%$ and $X=85\%$ for the simulated systems. Details of the experimental curing and determination of $X$ are included in the Supporting Information. We also note that no termination mechanism was implemented in these simulations. The extent of reaction is only constrained by hindered transport and steric trapping of radical-bearing species, both increasing as polymerization proceeds. 

One of the most notable physical transitions during polymerization of the liquid monomer is the gel point, loosely defined as when the propagating network spans the reaction volume. We estimate the gel point conversion ($X_{\text{gel}}$) as shown in Figure~\ref{fig:react}b, by tracking the size (monomer units) of the largest propagating polymer mass~\cite{hagbergExploringNetworkStructure2008} in the simulation. The inflection point of these curves indicates when the largest unit switches from rapidly growing by reaction with individual monomers to merging with other propagating chains. We use this inflection point to estimate $X_{\text{gel}}$. Experimental data directly corresponding to Figure~\ref{fig:react}b is not accessible, but we may validate our estimated $X_{\text{gel}}$ values by comparing with theoretical and experimental $X_{\text{gel}}$. The step-growth $X_{\text{gel}} = 50\%$ matches the Flory-Stockmayer (F-S) theoretical gel point.~\cite{floryMolecularSizeDistribution1941} Experimental studies of related thiol-ene systems report similarly precise agreement with the F-S model~\cite{chiouRealTimeFTIRSitu1997} and agreement has been reported in other simulated step-growth polymerization studies.~\cite{aramoonCoarseGrainedMolecularDynamics2016,estridgeSimulationStudiesDiblockCopolymer2015,estridgeEffectsCompetitivePrimary2018} The chain-growth system gels significantly earlier at 9\% conversion, similar to recent all-atom simulations of a different trifunctional acrylate.~\cite{karnesNetworkTopologyCrossLinked2020} This gel point is much lower than predicted by F-S theory, which only applies to step-growth systems. We validate the simulated triacrylate $X_{\text{gel}}$ by noting two experimental outcomes. First, the production of solid, photopolymerized pentaerythritol triacrylate with $X= 22\%$~\cite{oakdalePostprintUVCuring2016} indicates that $X_{\text{gel}} < 22\%$ for this triacrylate resin. Secondly we recently estimated $X_{\text{gel}} \approx 15\%$ for a diacrylate photoresin by real-time FTIR analysis. This suggests $X_{\text{gel}} < 15\%$ for related trifunctional acrylates (see Supporting Information for experimental details.)

In addition to reasonable agreement between simulated and experimental $X_{\text{gel}}$, Figure~\ref{fig:react}b confirms the general shape of chain- and step-growth thermoset polymers reported in all-atom simulations of both mechanistic classes.~\cite{karnesNetworkTopologyCrossLinked2020,estridgeSimulationStudiesDiblockCopolymer2015} This qualitative agreement suggests that our approach maintains the distinct network formation pathways of more specialized force fields and simulated reactivity.

\begin{figure}[h]
\includegraphics[width=5in]{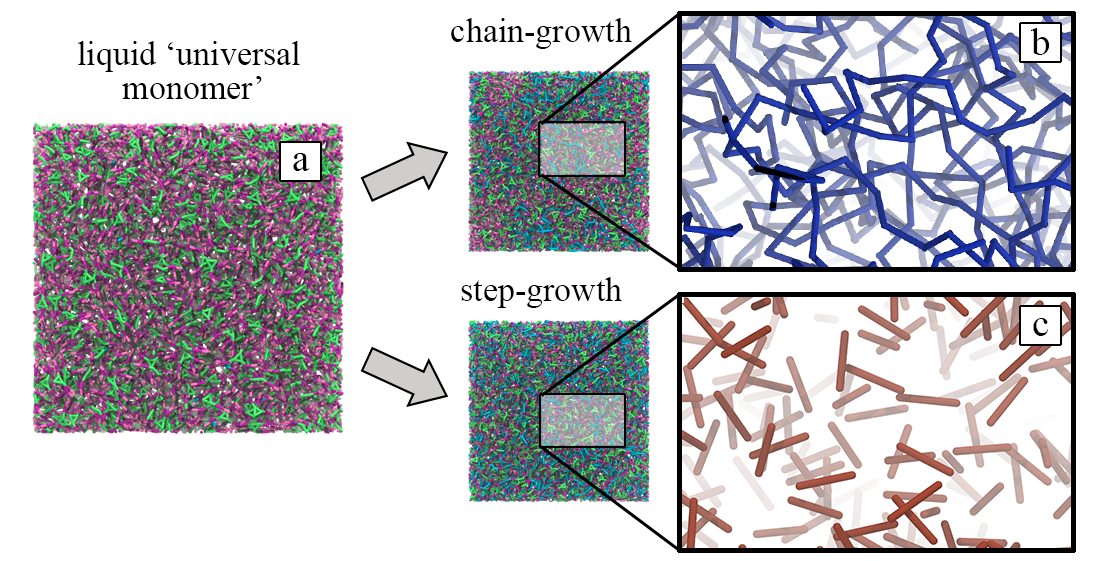}
\caption{\label{fig:morph} Snapshots of the liquid monomers and thermosets look similar. By hiding all nonreactive beads, only the monomer-monomer bonds are shown, revealing dramatic differences in the network topologies of the (a) chain-growth (b) and step-growth systems.}
\end{figure}

Molecular simulations also allow for detailed investigation of the evolving microscopic polymer networks. Figure \ref{fig:morph} shows simulation snapshots of the equilibrated liquid `universal monomer' (\ref{fig:morph}a) and the crosslinked thermosets resulting from application of each reaction `rule.' Differences between the post-reaction snapshots (center images, Figure \ref{fig:morph}) are not immediately apparent. Figures \ref{fig:morph}b and \ref{fig:morph}c zoom in on each thermoset and make all intra-monomer bonds invisible. This leaves only the inter-monomer bonds formed during the reactive CGMD simulations visible, revealing the respective network topologies. Presenting the simulations snapshots in this manner distinguishes the long, intertwining chains of acrylate sites in Figure \ref{fig:morph}b from from the isolated, individual thiol-ene linkages in Figure \ref{fig:morph}c.      

\begin{figure}[h]
\includegraphics[width=3.2in]{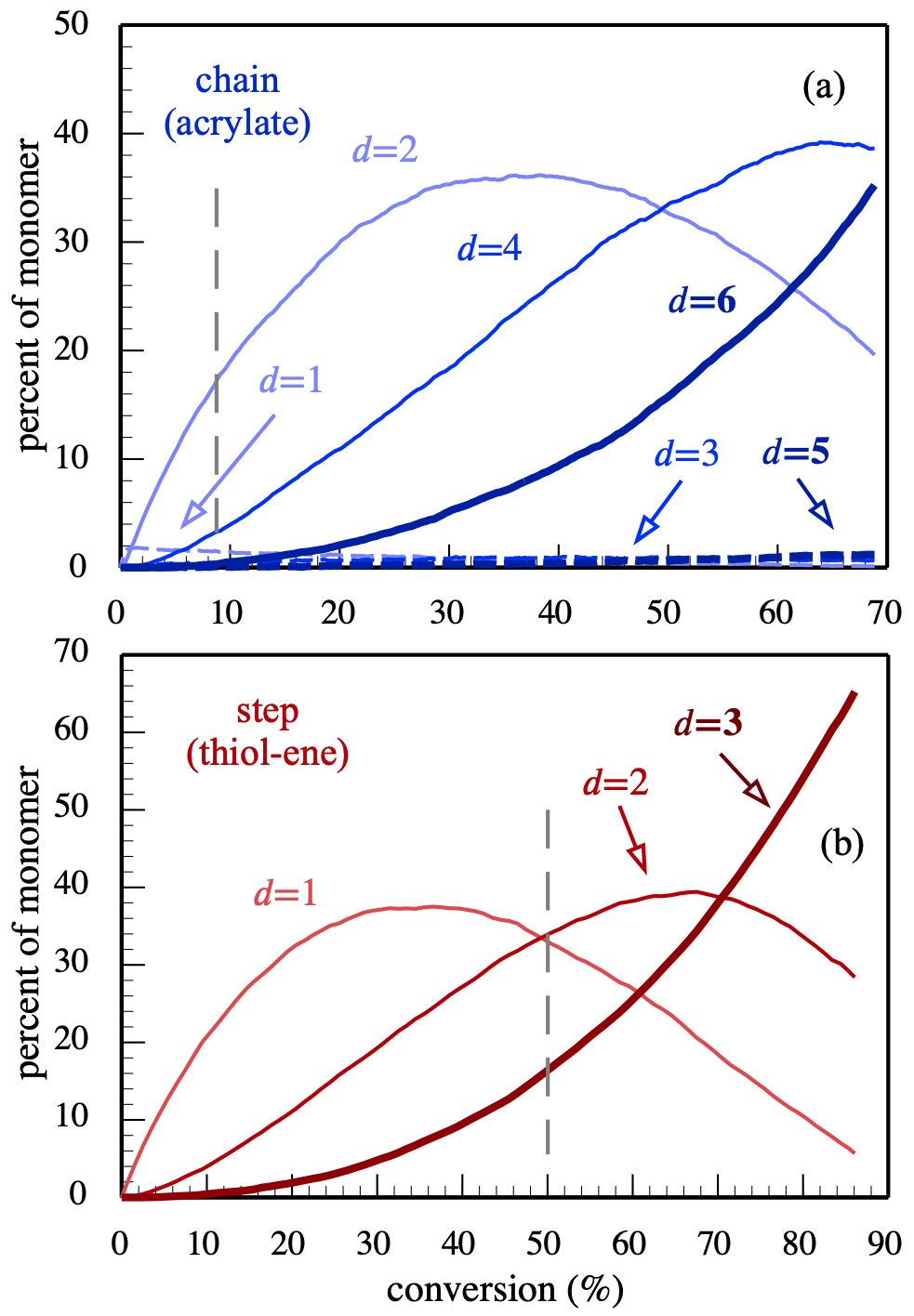}
\caption{\label{fig:deghist} Monomer connectivity histograms for the (a) chain-growth (acrylate) and (b) step-growth (thiol-ene) systems. Vertical dashed lines indicate the respective gel point conversions in each system.}
\end{figure}

The number of inter-monomer bonds that a each UM participates in (degree, $d$) is tracked and recorded as the simulation progresses. We remind the reader that each acrylate functional group may participate two inter-monomer bonds (be within a `chain') while thiol or ene functional groups may only participate in one, therefore triacrylate monomers have a maximum degree $d=6$ and trithiols and trienes a maximum $d=3$. Figure \ref{fig:deghist} shows histograms of monomer degree versus conversion for each reaction mechanism. These histograms agree well with previously reported simulations of triacrylate polymerization and compare reasonably well with simulations of epoxy thermosets.\cite{karnesNetworkTopologyCrossLinked2020,estridgeEffectsCompetitivePrimary2018} In Figure \ref{fig:deghist}a the $d=2$ curve corresponds to monomers participating in one linear acrylate chain, a trend that increases until approximately $X=35\%$, declining as crosslinking density increases and the $d=4$ and $d=6$ curves increase. Monomers of $d\geq3$ are most likely functioning as cross-linkers between multiple chains, except in the case where small loops or cycles are formed during the polymerization. In the step growth system the $d=1$ and $d=2$ curves appear to cross near the F-S gel point conversion of $X_\text{gel}=50\%$. This $d=1$ to $d=2$ crossover near the theoretical $X_\text{gel}$ was also observed in epoxy simulations for some formulations.~\cite{estridgeEffectsCompetitivePrimary2018} Dashed curves in Figure  \ref{fig:deghist}a indicate monomers with odd $d$. In the acrylate chain-growth simulations, odd degree is only possible in two cases: when a monomer has an acrylate site at the end of a propagating chain or in the case where a polymer has bonded to itself, a `self-cycle' in graph theory terminology.   

\begin{figure}[h]
\includegraphics[width=3.3in]{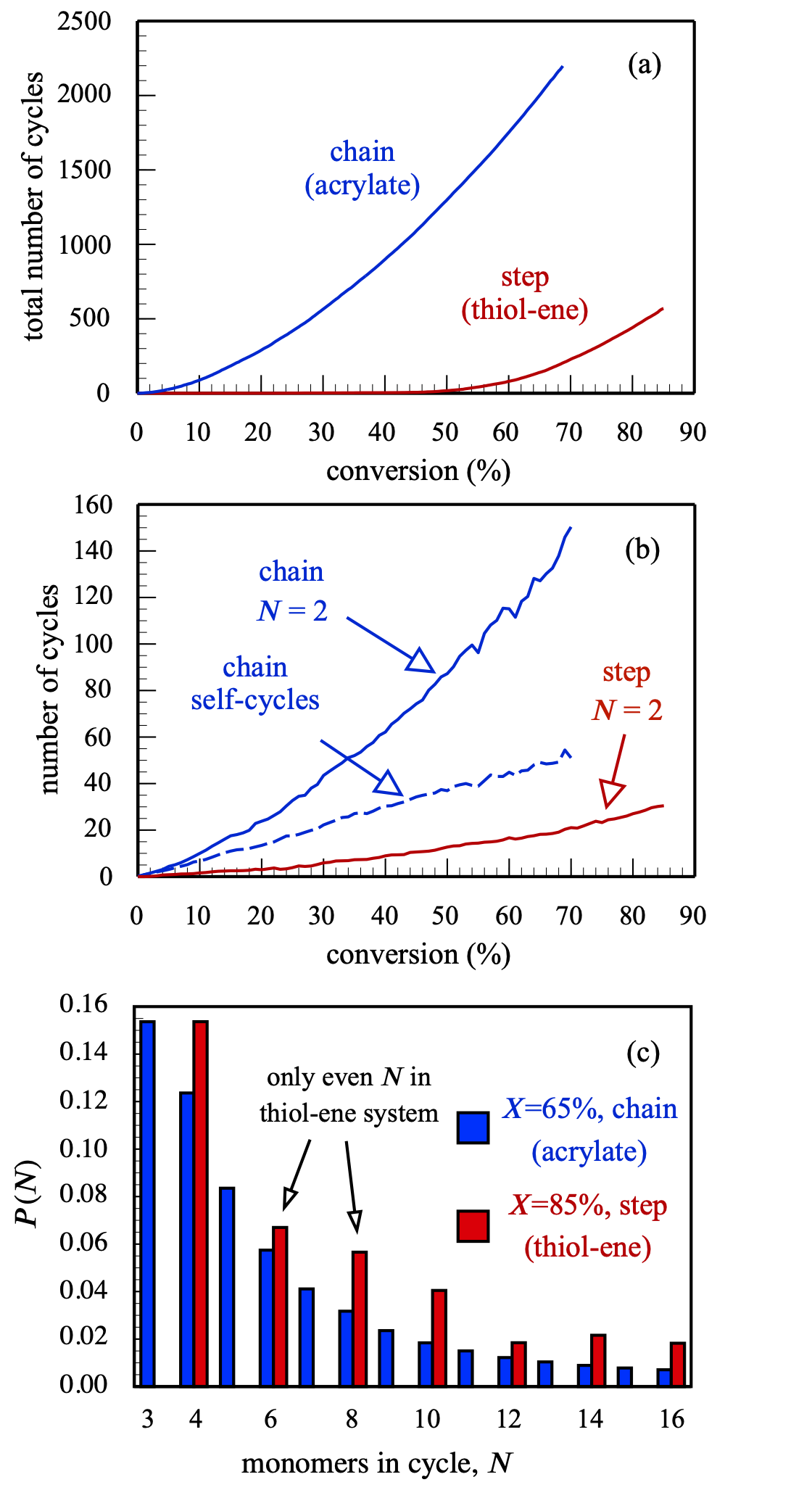}
\caption{\label{fig:cycles} (a) Total cyclomatic complexity of the chain- and step-growth systems as the networks evolve. (b) Subset of self- ($N=1$) and two-monomer ($N=2$) cycles in the systems and (c) cycle size histograms of the respective systems near their respective endpoint conversions.} 
\end{figure}

The presence of loops or cycles in the polymer network has interested researchers since the 1950s, features whose early appearance may delay the gel point and, at higher conversions, influence macroscopic properties of the resulting material.~\cite{langAnalysisGelPoint2020,korolevCalculationGelPoints1982,harrisRingFormationMolecular1955,monteferranteCapturingFreeRadicalPolymerization2022} Figure \ref{fig:cycles}a summarizes the evolution of cyclomatic complexity of the two thermosets. We note that these calculations consider only inter-monomer connectivity and ignore the 3-site isocyanurate centers of each monomer. Figure \ref{fig:cycles}a shows the total number of cycles in each system. The acrylate system noticeably forms cycles earlier and to a much greater number than in the thiol-ene system as polymerization progresses. These results may be expected due to the acrylates' network formation involving more inter-monomer linkages than thiol-ene and polymerization progressing by propagating strands of connected acrylate sites versus the thiol-ene's formation of spatially discrete, independent linkages. 

We also consider the size of cycles with the network, where $N$ is the number of monomers participating in a given cycle. Figure \ref{fig:cycles}b quantifies number of $N=1$ (self-) and $N=2$ cycles. The emergence of these small cycles is roughly linear with conversion with the exception of the acrylate's $N=2$, which suggests that their formation is not a strong function of local topology or steric conditions imposed by the surrounding network. Self-cycles are not permitted in the thiol-ene system, since neither monomer can bond with itself. Additionally, all cycles in the thiol system must be of even $N$. 

Despite the differing number of total cycles and limitations on $N$, the cycle size distributions of the two thermosets look similar. Figure \ref{fig:cycles}c shows a probability histogram of cycle sizes for the two systems near their respective endpoint conversions, $X=65\%$ for the acrylate and $X=85\%$ for thiol-ene. 

\clearpage

\begin{figure}[h]
\includegraphics[width=3.3in]{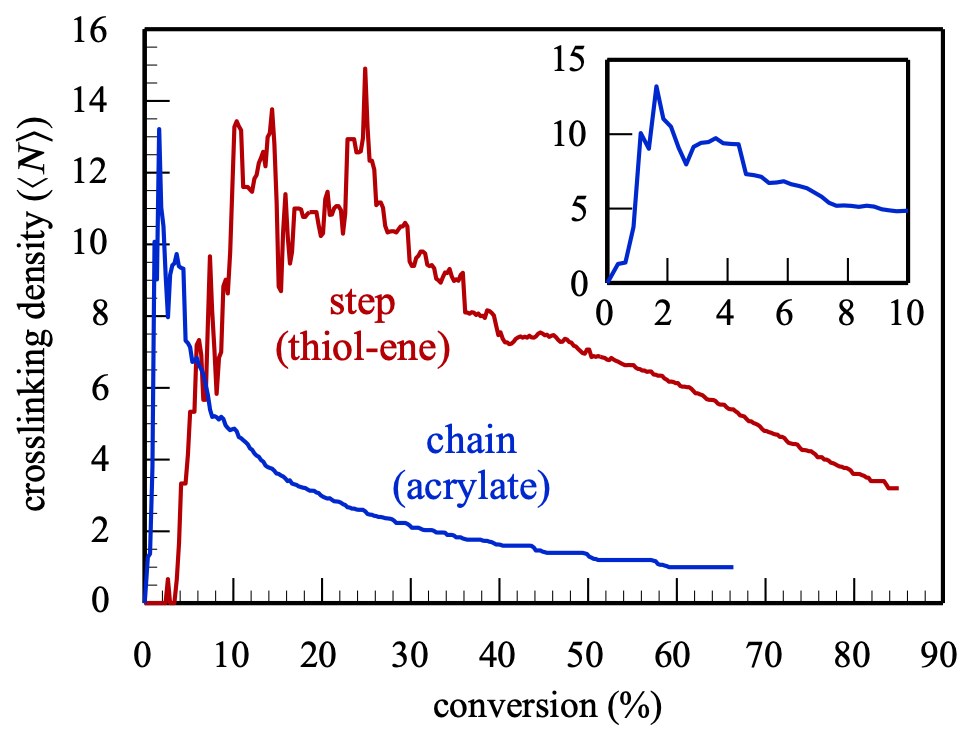}
\caption{\label{fig:xl} Average crosslinking density of the largest connected component in the chain- and step-growth systems. Inset: Acrylate crosslink density for $0<X<10\%$.}
\end{figure}

Figure \ref{fig:xl} shows the average crosslinking densities as a function of conversion, where the units of the $y$-axis is the average number of monomer units between each crosslink, $\langle N \rangle$. One obvious feature of these curves is the noise in each prior to the respective gel point conversions. The major contribution to this noise is that this calculation considers the crosslinking density of the largest connected component in the simulation. The identity of the largest connected component may change as the reaction progresses prior to $X_{\text{gel}}$. The rapid initial increase in the curves indicates linear growth of the components, followed by a gradual decline. Notably, the acrylate system appears to asymptotically approach a final value of $\langle N \rangle \approx 1$ while the thiol-ene crosslinking density continues to increase linearly with conversion until the polymerization stops progressing.

\begin{figure}[h]
\includegraphics[width =6in]{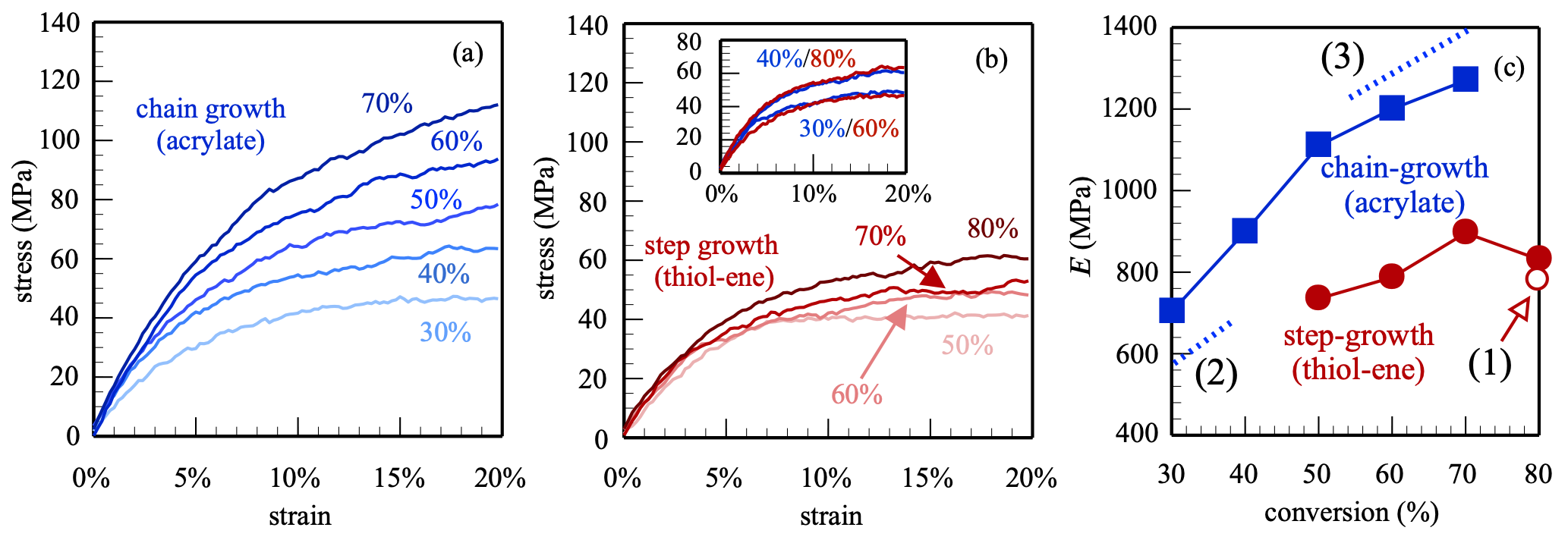}
\caption{\label{fig:strstr} Stress-strain curves for (a) chain-growth and (b) step-growth systems at selected conversions ($X$). The inset in (b) overlays selected step-growth (red) and chain-growth (blue) curves. Color-coded percentage labels indicate $X$ of the respective thermoset. (c) Young’s modulus data for the simulated thermosets.}
\end{figure}

Lastly, we consider the emergent mechanical properties of the thermosets. Figure~\ref{fig:strstr} shows stress-strain data obtained from simulated uniaxial extension at selected $X$. Figure~\ref{fig:strstr}a and b show stress-strain curves for chain- and step-growth simulations. In chain-growth, the stress-strain response increases with conversion and is significantly greater than the respective step-growth systems. We attribute most of the differences between respective step and chain configurations to one main factor: Fewer new covalent bonds are formed with increasing conversion in the step-growth system. As outlined earlier, step-growth conversions of 60\% and 80\% represent the same number of new inter-monomer bonds as 30\% and 40\% conversion in the chain-growth system and, despite the dramatic differences in microscopic morphology, they display similar stress-strain response. The inset of Figure~\ref{fig:strstr}b shows these four curves together. This indicates that the number of inter-monomer linkages, not the morphology of the thermoset network, is the most important factor in these tensile strength simulations. 

Figure~\ref{fig:strstr}c shows the Young’s modulus ($E$) extracted from each stress-strain simulation (solid symbols). Both systems show an increase in $E$ with increasing $X$, more pronounced in the acrylate system. We ascribe this difference to greater inhomogeneity during network evolution, with the step-growth mechanism proceeding by the formation of many small clusters that crosslink later in the polymerization (see Figure~\ref{fig:react}b). Less significant topological evolution, coupled with fewer covalent linkages being formed, results in the differences in slope. Simulated thiol-ene moduli agree well with the previously reported value of $782 \pm 19$ MPa for the same isocyanurate system at endpoint conversion, labeled (1) in Figure~\ref{fig:strstr}c.~\cite{mongkhontreeratUVInitiatedThiol2013} The simulated trifunctional acrylate system also agrees fairly well with experimental data sets reported for similar trifunctional acrylates, indicated by the dotted lines in Figure~\ref{fig:strstr}c. These include reported $E$ vs. $X$ for the lower conversions in this work, approximately 580 to 640 MPa for conversions of 32-40\% (labeled (2))~\cite{jiangTwophotonPolymerizationInvestigation2014} and approximately 1200-1400 MPa for conversions of 55-70\% (labeled (3)).~\cite{cichaYoungModulusMeasurement2012,cichaEvaluation3DStructures2011} We note that these authors also obtained higher values of $E$ ($\sim$1400-1900 MPa) by varying photoinitiator and processing conditions and we compare our results with the experiment that more closely resembles our simulation conditions.~\cite{cichaYoungModulusMeasurement2012,cichaEvaluation3DStructures2011} Section S3 of the Supporting Information contains more details regarding previously reported experimental data. We reserve the \textit{in silico} exploration of spatiotemporal radical initiation versus $E$ for future studies with larger simulation cells.


\section{Conclusions}

The ‘universal monomer’ approach toward simulation of thermoset polymers or other materials allows for a rapid survey of chemical mechanism space and isolates the reaction mechanism’s effect on the resulting properties, both micro- and macroscopic. Extension of this concept to other polymerization mechanisms (e.g., epoxy, living, etc.) is straightforward. These comparisons of step and chain-growth polymerization show the expected network formation and simulated thermomechanical testing suggests that the number of newly formed covalent linkages, not only the microscopic morphology of the polymer network dictates the mechanical properties within a range of experimentally accessible conversions. Our proof-of-concept work found the differences in $X_{gel}$ and tensile testing results are well captured by the UM approach and therefore strong functions of chemical reaction mechanism.
Subsequent work will expand the concept of the UM to include more reaction mechanisms and the copolymerization of tri- and bifunctional monomers.

\begin{acknowledgement}
The authors thank J. R. Gissinger, J. J. Schwartz, T. M. Bryson, W. F. D. Bennett, T. S. Carpenter, and H. I. Ingólfsson for many insightful conversations. This work was supported by Lawrence Livermore National Laboratory’s Laboratory-Directed Research and Development (LDRD) funding, project 19-ERD-012 and performed under the auspices of the U.S. Department of Energy by Lawrence Livermore National Laboratory under Contract DE-AC52-07NA27344, release number LLNL-JRNL-823655.

\end{acknowledgement}

\begin{suppinfo}

Additional parameterization and MD simulation details and summaries of experimental thermoset preparations, determination of $X$, and estimation of  $X_{\text{gel}}$ are included in the Supporting Information.

\end{suppinfo}

\bibliography{betterBibTex}

\end{document}


\maketitle

\section{Simulation Details}
\subsection{Customizing the Martini Model}

\begin{figure}[ht!]
\centering
\includegraphics[scale=1.3]{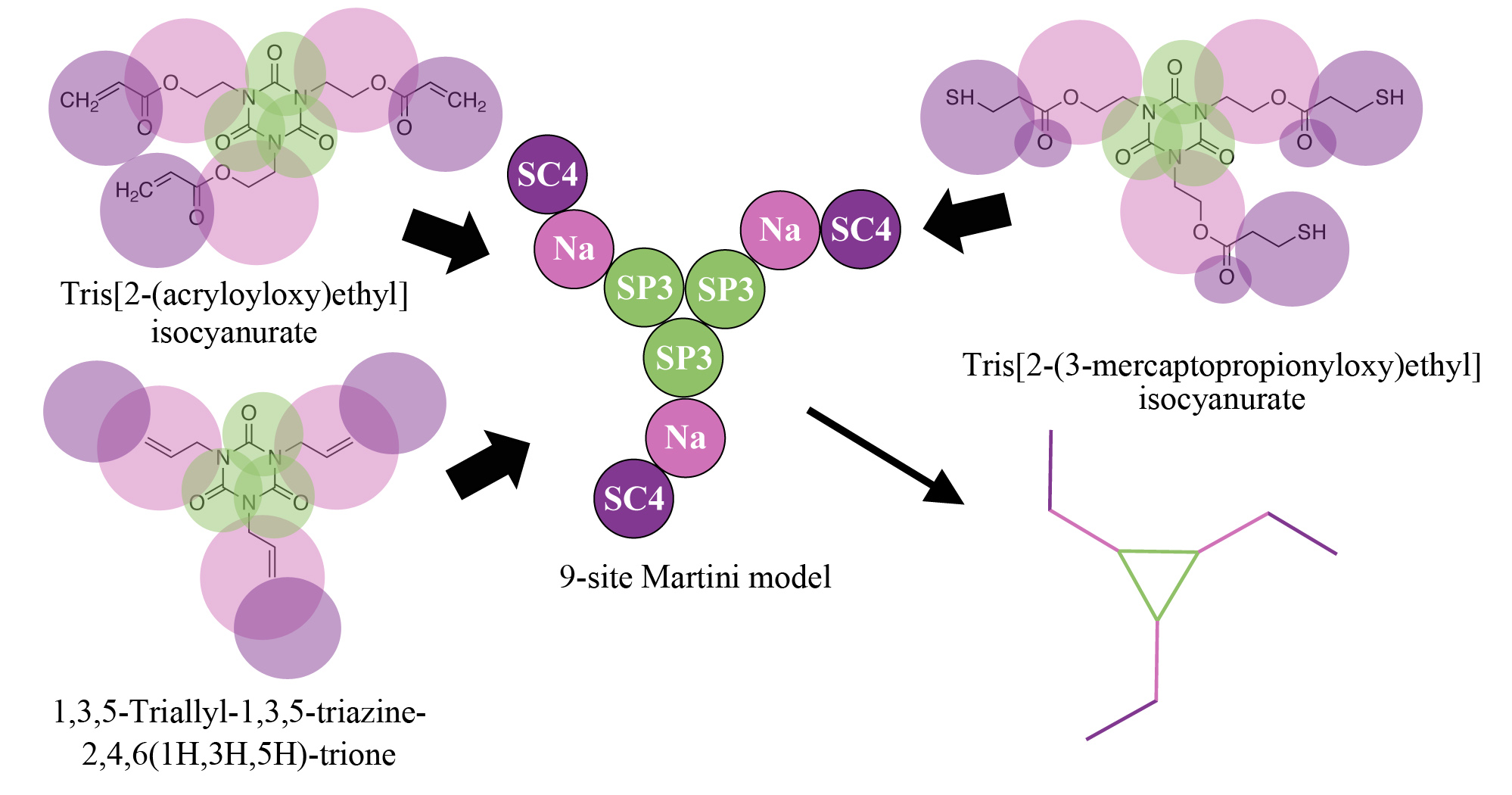}
\caption{One Martini model (center) represents all three trifunctional monomers. The green, pink, and purple regions define the atom centers composing each bead.}
\label{fig:mono2tini}
\end{figure}

Individual atoms were assigned to Martini beads~\cite{marrinkMARTINIForceField2007} as shown in Figure~\ref{fig:mono2tini}. 


\clearpage

\begin{figure}[ht!]
\centering
\includegraphics[scale=0.7]{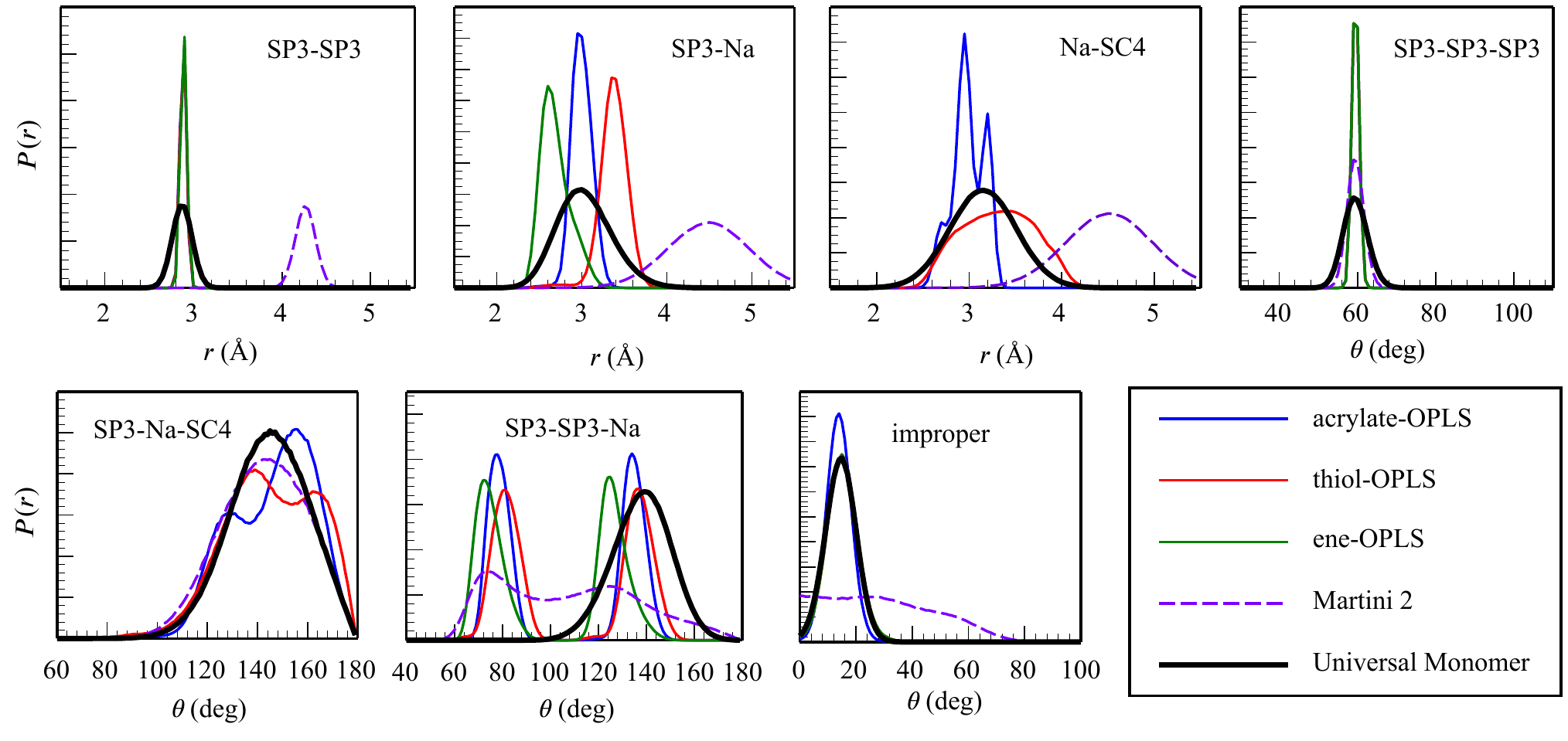}
\caption{Probability distributions summarizing the intramolecular geometries of all-atom simulations mapped onto CG models (blue, red, and green curves) and guide reparameterization of the default Martini 2 force field (dashed purple curves). Several iterations result in the `universal monomer' model, which tries to best approximate all three OPLS-based models (bold black curves).}
\label{fig:param}
\end{figure}

Using in-house code, we calculated coarse-grained bonds, bends, and SP3-SP3-SP3-Na improper dihedrals based on the centers of mass of each bead grouping. These CG-mapped geometries were collected for each molecule in each saved configuration and assembled into probability distributions representing each component of the coarse-grained force field topology. Figure~\ref{fig:param} shows the CG-mapped distributions for the three all-atom simulations are shown as the red, blue, and green curves. The universal monomer (UM) paramaterized for this work is represented by the bold black curves and the Martini 2 force field, included for reference, is shown as dashed purple curves.

\subsection{Equilibration}
\begin{figure}[ht!]
\centering
\includegraphics[scale=0.7]{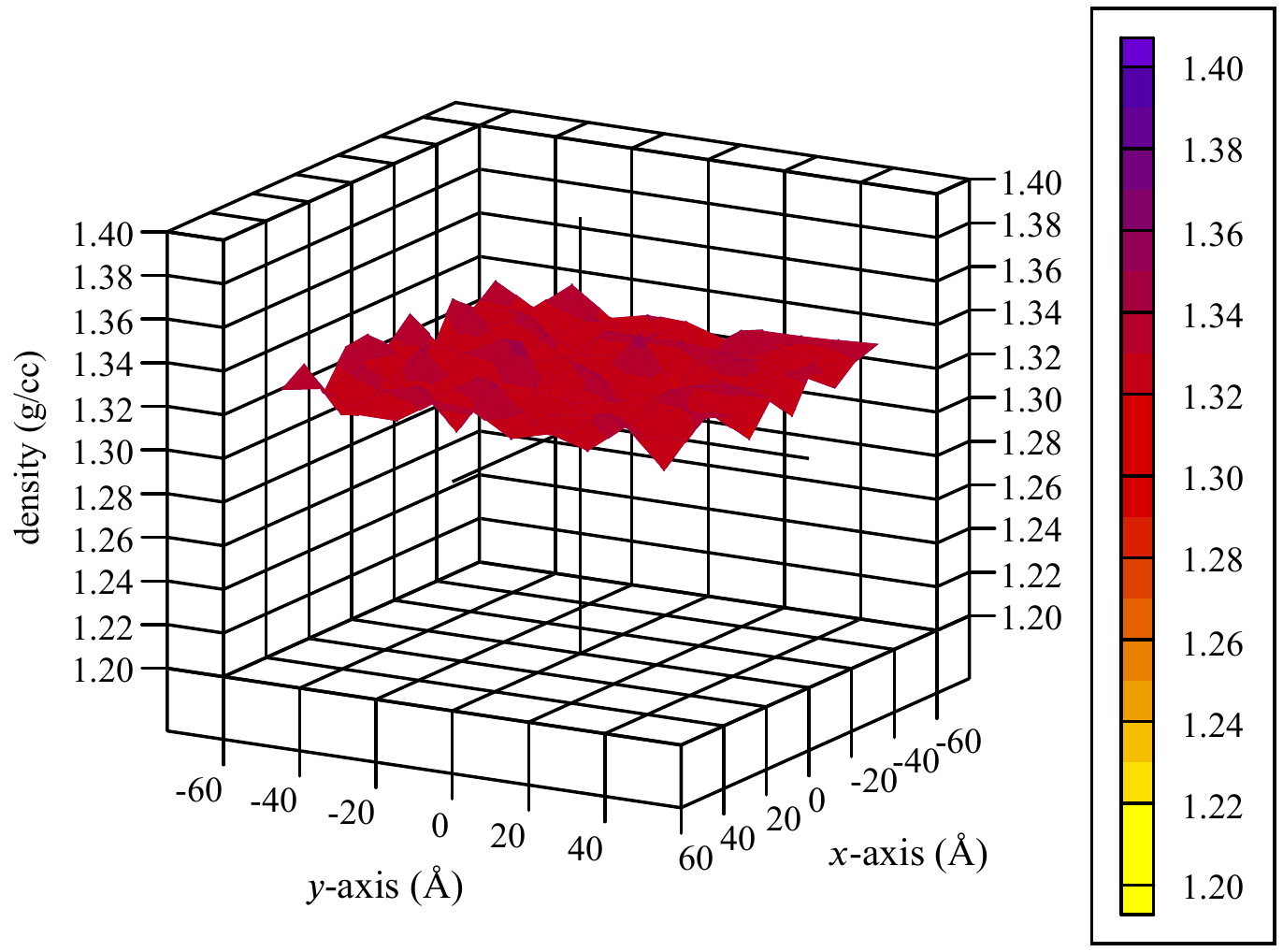}
\caption{Representative density profiles collected during the final equilibration step.}
\label{fig:densProf}
\end{figure}

Reactive production runs consisted of 2000 9-site `universal monomers' in a simulation box with periodic boundaries in the $x$, $y$, and $z$ directions. To prepare the starting configurations for these simulations, MD runs with short ($\mathit{dt} <$ \SI{1}{fs}) timesteps were used to relax any high-energy configurations followed by longer runs with increasingly larger time steps. The final equilibration step consisted of a 20 ns trajectory performed in the NPT ensemble at \SI{450}{K} and \SI{1}{atm} with \SI{10}{fs} time steps. Figure~\ref{fig:densProf} shows a representative density profile, displayed as projections that integrate over the $x$-axis, of the equilibrated liquid monomer box. For reference, simulation boxes of the neat OPLS isocyanurate monomers have densities of 1.15 (-ene), 1.33 (-thiol), and 1.25 g/cm$^3$ (-acrylate). We prepared six independent UM starting configurations for each of the two polymerization mechanisms, radical chain-growth and radical step-growth.

\subsection{Reactive Scheme}
\begin{figure}[ht!]
\centering
\frame{\includegraphics[scale=0.75]{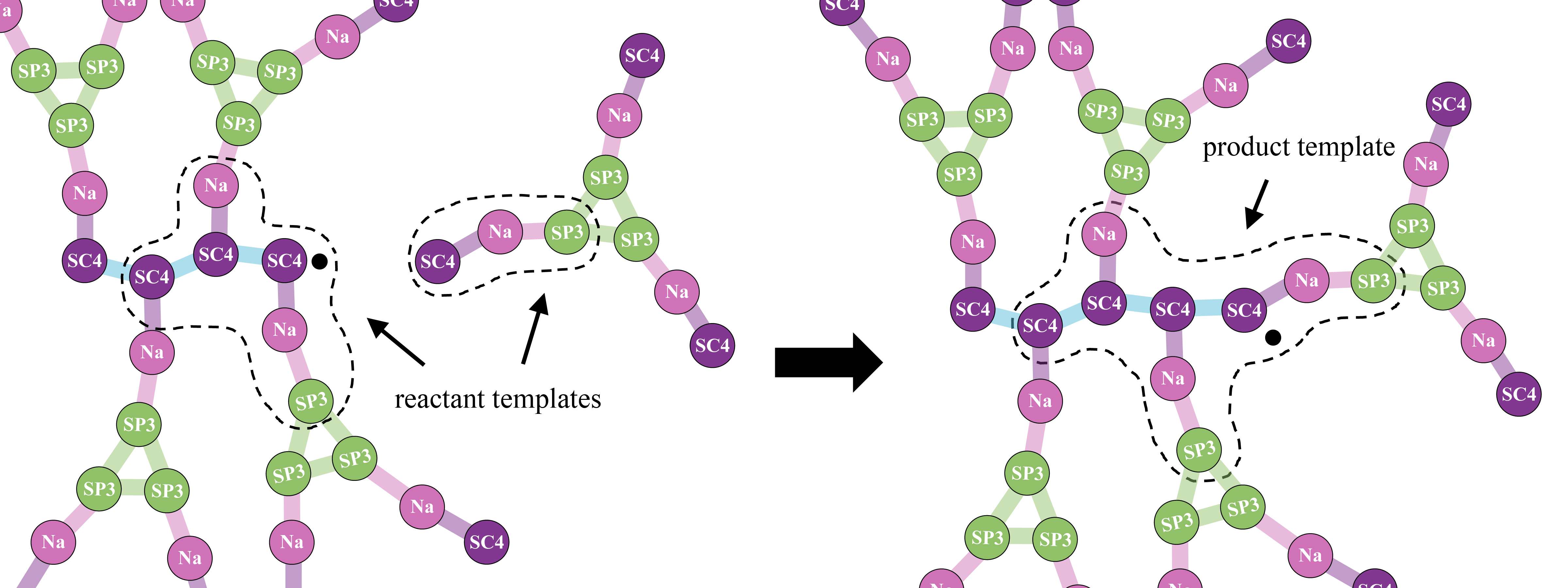}}
\caption{Representative reactive scheme for radical chain-growth polymerization of the universal monomer. The beads within the dashed lines belong to reactant templates (left-hand side of cartoon) or product template (right-hand side).}
\label{fig:scheme}
\end{figure}

We use the REACTER protocol, a distance cutoff-based reactive scheme to simulate polymerization in our CGMD trajectories~\cite{gissingerModelingChemicalReactions2017}. A new bond is formed between two potential reactive sites if the distance between them is smaller than some threshold distance $r$. In this work $r$ is chosen to be 4.7 \AA. This same cutoff is used for radical transfer in the thiol-ene scheme: a `radical -ene' within $r$ of an unreacted thiol bead may transfer the radical to the thiol. There is no restriction, aside from steric limitations, prohibiting intra-molecular reaction or radical transfer. When a new bond if formed, the affected local force field topology is updated accordingly, e.g. $i$-$j$-$k$ bends that bridge the new bond are added. Figure~\ref{fig:scheme} contains a cartoon schematic of our implementation. At time-step zero, 20 reactive sites (acrylate or thiol) are mutated to radical-bearing sites, which corresponds to 1\% of the monomers in each simulation box. 

\subsection{Uniaxial Stress-Strain Response} 
Starting configurations for simulated stress-strain response were obtained similarly to the $T_g$ simulations described above. The simulation box was elongated in the $x$-dimension at an engineering strain rate of \SI{1e-6}{\per\second} while the $y$- and $z$-dimensions were controlled by the Nos\'e--Hoover barostat with a pressure set point of \SI{1}{atm} and damping parameter of \SI{10}{ps}. Stress data from the molecular dynamics trajectories was binned into 100 equally-spaced intervals between 0 and 20\% engineering strain. Data reported in the main text represents the average of 6 independent uniaxial extension simulations.  

\section{Experimental Methods}
\subsection{Resin formulation and sample preparation}
Tris[2-(acryloyloxy)ethyl] isocyanurate (TAE-ICN), 1,3,5-Triallyl-1,3,5-triazine- 2,4,6-(1H,3H,5H)-trione (TA-ICN), and photoinitiator 2-Methyl-4'-(methylthio)-2-morpholinopropiophenone or Irgacure 907 were purchased from Sigma Aldrich. Tris[2-(3-mercaptopropionyloxy)ethyl] isocyanurate (TME-ICN) was purchased from TCI America. All chemicals were used without further purification. Step growth formulations used a stoichiometric ratio of TA-ICN to TME-ICN, while chain growth formulations used TAE-ICN similarly outlined in Cook et al.\cite{cookHighlyTunableThiolEne2020} The photosensitive resin prepared in batches varying in photoinitiator concentrations was cast into an ASTM D638 Type V dogbone specimen mold. UV exposure times and intensities used to attain the range of functional group conversions are detailed in Table ~\ref{table:resins}. All samples were flood cured with a \SI{405}{nm} XYZprinting UV curing chamber, aside from the 50\% conversion step growth sample that was cured with a \SI{366}{nm} Model UVGL-25 Mineralight Lamp. While the efficient ``click''-chemistry nature of the thiol-ene reaction motivates its use in several applications, this same efficiency makes cessation of the reaction at intermediate conversions between the onset of gelation and final cure challenging. As a result, our experimental survey was only able to capture experimental thiol-ene data points near the onset of gelation and at the endpoint. 

\begin{table}[ht!]
\centering
\begin{tabular}{ c M{2cm} M{2.2cm} M{2.2cm} M{1.8cm}}
 \hline
 sample & conversion (\%) & photoinitiator (mM) & light intensity (mw cm$^{2}$) & exposure time (min)\\
 \hline
 acrylate & 50\% & 10 & 18 & 1\\
 acrylate & 60\% & 10 & 18 & 5\\
 \textbf{acrylate} & \textbf{66\%} & \textbf{40} & \textbf{18} & \textbf{20}\\
 thiol-ene & 50\% & 0.5 & 2 & 4\\
 \textbf{thiol-ene} & \textbf{89\%} & \textbf{10} & \textbf{18} & \textbf{5}\\
 \hline
 \end{tabular}
 \caption{Resin formulation and sample exposure conditions. Bold entries approximate endpoint conversions of the respective systems.}
\label{table:resins}
\end{table}

\subsection{Chemical Conversion}
Functional group conversions ($X$) were measured with a custom backscattered Raman spectrometer. For step growth systems, the thiol peak at \SI{2555}{\per\cm} and alkene peak at \SI{1650}{\per\cm} areas were monitored relative to the unchanging \SI{1244}{\per\cm} of a tertiary aromatic amine signal.~\cite{yangCharacterizationWelldefinedPoly2011} Analysis for chain growth systems focused on the acrylate \SI{1640}{\per\cm} peak proportionate to the full spectrum area. Conversion was calculated by measuring the change in signal peak area and the normalization area

\begin{equation}
    X (\%) = \left[ 1- \left( \frac{A_{\text{peak}}/A_{\text{norm}}}{A^0_{\text{peak}}/{A^0_{\text{norm}}}} \right) \right] \times 100\%
\end{equation}

\medskip
\medskip
\noindent where $A^{0}$ represents the area from the uncured resin.~\cite{oakdalePostprintUVCuring2016} 

\subsection{Estimated Chain-Growth Gel Point Conversion}
The chain-growth gel point conversion was estimated experimentally with a 75:25 wt\% mixture of bisphenol A glycerolate (1glycerol/phenol) diacrylate (BPAGDA) to poly)ethylene glycol diacrylate (PEGDA, Mn = \SI{250}{\g\per\mol}) and 40mM 2-methyl-4'-(methylthio)-2-morpholinopropiophenone or Irgacure 907. All materials were purchased from Sigma Aldrich without further purification. UV Fourier Transform IR (FTIR, Bruker Vertex 80) spectroscopy and UV rheology (TA Instruments DHR-1 with the UV attachment, 0.0012 strain, \SI{3}{\rad\per\s}) data was acquired and overlapped to determine the acrylate conversion ($X$) at \SI{810}{\per\cm} when the loss modulus equals the storage modulus, indicating the point of gelation.~\cite{chiouRealTimeFTIRSitu1997,cookHighlyTunableThiolEne2020} A \SI{25}{\mW\per\cm^2} \SI{405}{\nm} light intensity (OmniCure S2000, Excelitas, \SI{405}{nm} Thorlabs band-pass filter) was kept consistent for both experiments. Due to variance in the induction period in the UV FTIR and UV rheology experiments, the initiation to propagation kinetic shoulder was adjusted to line up, thus providing an estimate for the gel point conversion around 15\%.

\begin{figure}[ht!]
\centering
\includegraphics[scale=0.8]{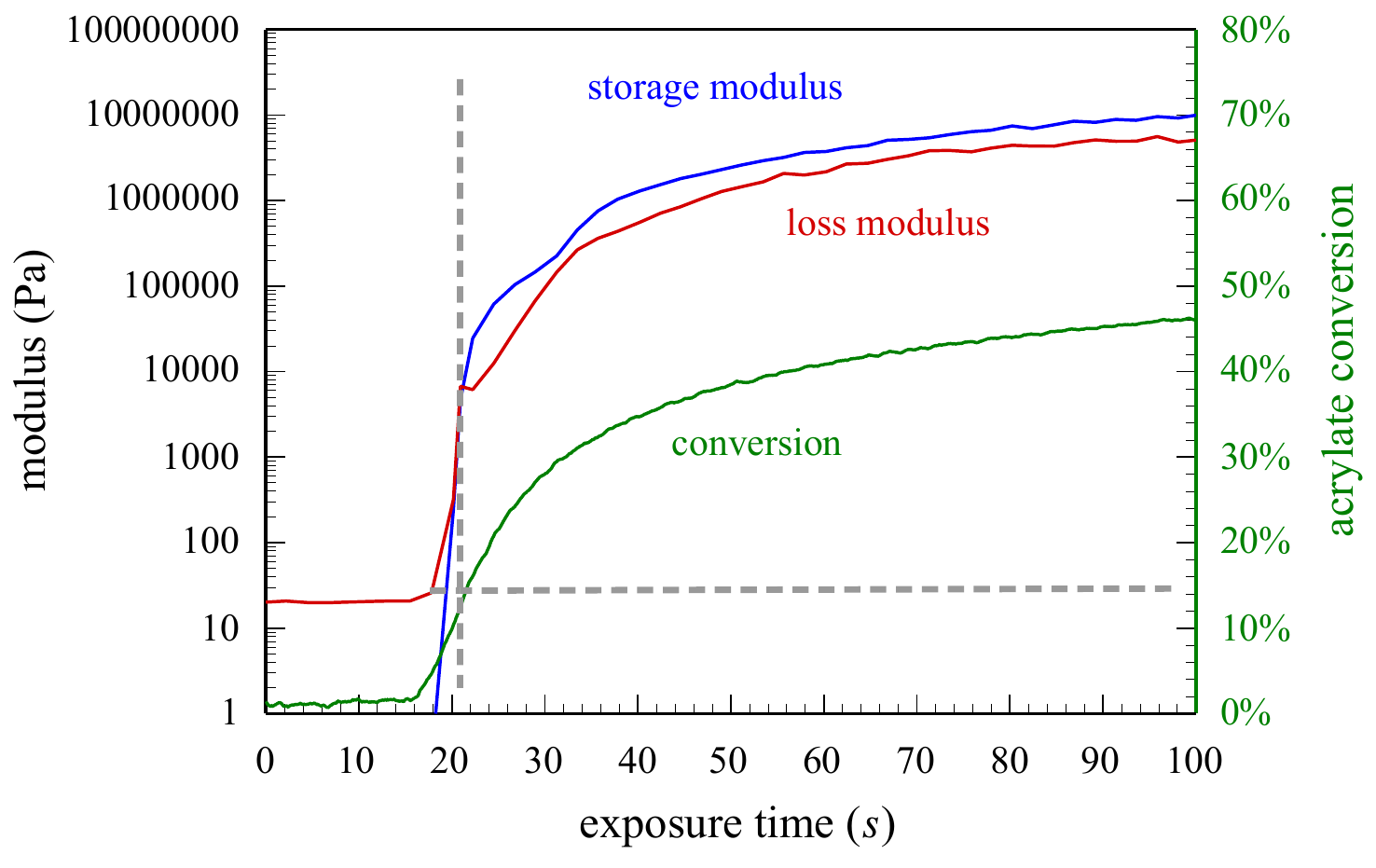}
\caption{Chain growth gel point conversion estimation when comparing the gelation time found in UV rheology with the acrylate conversion found in FTIR.}
\label{fig:gelPoint}
\end{figure}

\section{Tensile testing data}
\subsection{Experimental Tensile testing data}
Experimental tensile testing data was collected by Jiang et. al. for the IP-L 780 (Nanoscribe GmbH) photoresist and reported in Reference \cite{jiangTwophotonPolymerizationInvestigation2014} as reduced Young's Modulus, $E_r$, which we may convert to conventional $E$ by the relationship

\begin{equation}
    \frac{1}{E_r}=\frac{(1-\nu^2)}{E}+\frac{(1-\nu_i^2)}{E_i},
\end{equation}

\noindent noting that the equation in Reference \cite{jiangTwophotonPolymerizationInvestigation2014} contains a typographical error: the exponent on the Poisson ratio of the resin sample, $\nu$, was omitted. The authors provide values of 0.02 for $\nu_i$ and 1140 GPa for $E_i$ but do not provide $\nu$. We use $\nu=0.3$ for our conversion, a conservative choice not critical in the context of analyzing our results. A change to $\nu=0.4$ shifts the range of $E$ by less than 100 MPa. Higher conversion tensile data estimates were compiled from works of Cicha et. al. that study a blended trifunctional acrylate photoresin consisting of a 1:1 weight percent mixture of ethoxylated (20/3)-trimethylolpropanetriacrylate (ETA, Sartomer 415) and trimethylolpropane triacrylate (TTA, Genomer 1330).~\cite{cichaEvaluation3DStructures2011,cichaYoungModulusMeasurement2012} In these works, the authors present laser power versus conversion and laser power versus Young's modulus results for two formulations. These formulations differ only by the photoinitiator added to the resin, either B3FL or E,E-1,4-bis[40-(N,N-di-n-butylamino) styryl]-2,5-dimethoxy- benzene (R1), both added at the same concentration of $6.3 \times 10^{-6}$ mol photoinitiator/g resin. 

\section{Acknowledgements}This work was supported by Lawrence Livermore National Laboratory’s Laboratory-Directed Research and Development (LDRD) funding, project 19-ERD-012 and performed under the auspices of the U.S. Department of Energy by Lawrence Livermore National Laboratory under Contract DE-AC52-07NA27344, release number LLNL-JRNL-823655.

\printbibliography